\documentclass[fleqn,10pt]{wlscirep}
\usepackage[utf8]{inputenc}
\usepackage[T1]{fontenc}
\usepackage{listings}
\usepackage{tabularx}
\usepackage{booktabs}
\usepackage{listings}
\usepackage{xcolor}
\usepackage{changepage}
\title{
Exploring the Efficacy of Base Data Augmentation Methods in Deep Learning-Based Radiograph Classification of Knee Joint Osteoarthritis}

\author[1,2,*]{Fabi Prezja}
\affil[1]{University of Jyväskylä, Faculty of Information Technology, Jyväskylä, Finland}
\affil[2]{Finnish Artificial Intelligence Research Network, Jyväskylä, Finland}

\author[3,4]{Leevi Annala}
\affil[3]{University of Helsinki, Faculty of Science, Department of Computer Science, Helsinki, Finland}
\affil[4]{University of Helsinki, Faculty of Agriculture and Forestry, Department of Food and Nutrition, Helsinki, Finland}

\author[1]{Sampsa Kiiskinen}
\affil[1]{University of Jyväskylä, Faculty of Information Technology, Jyväskylä, Finland}

\author[1]{Timo Ojala}
\affil[*]{faprezja@jyu.fi}

\begin{abstract}
Diagnosing knee joint osteoarthritis (KOA), a major cause of disability worldwide, is challenging due to subtle radiographic indicators and the varied progression of the disease. Using deep learning for KOA diagnosis requires broad, comprehensive datasets. However, obtaining these datasets poses significant challenges due to patient privacy concerns and data collection restrictions. Additive data augmentation, which enhances data variability, emerges as a promising solution. Yet, it's unclear which augmentation techniques are most effective for KOA. This study explored various data augmentation methods, including adversarial augmentations, and their impact on KOA classification model performance. While some techniques improved performance, others commonly used underperformed. We identified potential confounding regions within the images using adversarial augmentation. This was evidenced by our models' ability to classify KL0 and KL4 grades accurately, with the knee joint omitted. This observation suggested a model bias, which might leverage unrelated features for classification currently present in radiographs. Interestingly, removing the knee joint also led to an unexpected improvement in KL1 classification accuracy. To better visualize these paradoxical effects, we employed Grad-CAM, highlighting the associated regions. Our study underscores the need for careful technique selection for improved model performance and identifying and managing potential confounding regions in radiographic KOA deep learning.
\end{abstract}
\begin{document}

\flushbottom
\maketitle
%
%
\thispagestyle{empty}

\section*{Introduction}

The past decade has seen a considerable surge in the integration of artificial intelligence into medicine\cite{wang2019deep,beam2018big}, riding on the wave of the dramatic growth in deep machine learning methods\cite{lecun2015deep}. Medicine has emerged as a crucial field for applying these advanced technologies, with deep learning primarily targeting clinical decision support and data analysis. These systems, adept at examining medical data to discover patterns and relationships, span a diverse range of applications. They have demonstrated significant progress in predicting patient outcomes \cite{kather2019predicting,courtiol2019deep,diamant2019deep}, as well as enhancing diagnostics and disease classification \cite{esteva2017dermatologist,han2017breast,bakator2018deep,prezja2023improved,prezja2023improving}. Beyond analysis and classification, deep learning has proven effective in data segmentation \cite{isensee2021nnu,liu2021review} and has even made strides in the generation \cite{chuquicusma2018fool,calimeri2017biomedical,frid2018gan,thambawita2021deepfake,annala2020generating} and anonymization of medical data\cite{shin2018medical,yoon2020anonymization,torfi2022differentially,kasthurirathne2021generative,prezja2022deepfake,prezja2022synthetic}. The utility of these advances in the field of Osteoarthritis (OA), however, presents its unique set of challenges.

OA, with knee joint osteoarthritis\cite{yeoh2021emergence,saarakkala2010depth, laasanen2003biomechanical} (KOA) being especially prevalent\cite{HUNTER20191745}, is a primary global cause of disability\cite{hermans2012productivity}, with estimated expenditures reaching up to 2.5\% of the Gross National Product in western countries\cite{hermans2012productivity}. Its early detection is often impeded by subtle radiographic markers and disease progression variability\cite{HUNTER20191745,yeoh2021emergence}. Leveraging deep learning for diagnosing KOA \cite{tiulpin2019multimodal,tiulpin2018automatic,tiulpin2020automatic} depends heavily on the availability of diverse and extensive data sets. However, obtaining such data sets is problematic, constrained by patient privacy considerations\cite{centers2003hipaa,voigt2017eu}, data collection restrictions, and the inherent progression of OA. Various studies have utilized data augmentation techniques as a workaround, creating artificial data variability. Specifically for KOA, two leading data augmentation methods are employed: affine and online, where random transformations occur during training, and additive and offline, manipulating the base (original) data prior to training to generate more data points. These techniques, often used in tandem, have proven successful in enhancing performance and mitigating overfitting. However, until now, there has been no systematic exploration to determine which technique is most effective for the task at hand, nor which ones might be less beneficial. In addition, no prior research has probed the realm of adversarial augmentation, a tactic designed to deceive the underlying system into delivering high performance while excluding or distorting essential radiograph characteristics. This approach could potentially identify confounding regions within images, thereby enhancing validation processes.

In this study, we address these research gaps. Our focus centers on discerning the most suitable base augmentation technique for the task at hand and pinpointing potential confounding regions present within the radiographs (with adversarial augmentation).

\section*{Materials and Methods}

\begin{figure}[!ht]
\centering
\includegraphics[scale=0.45]{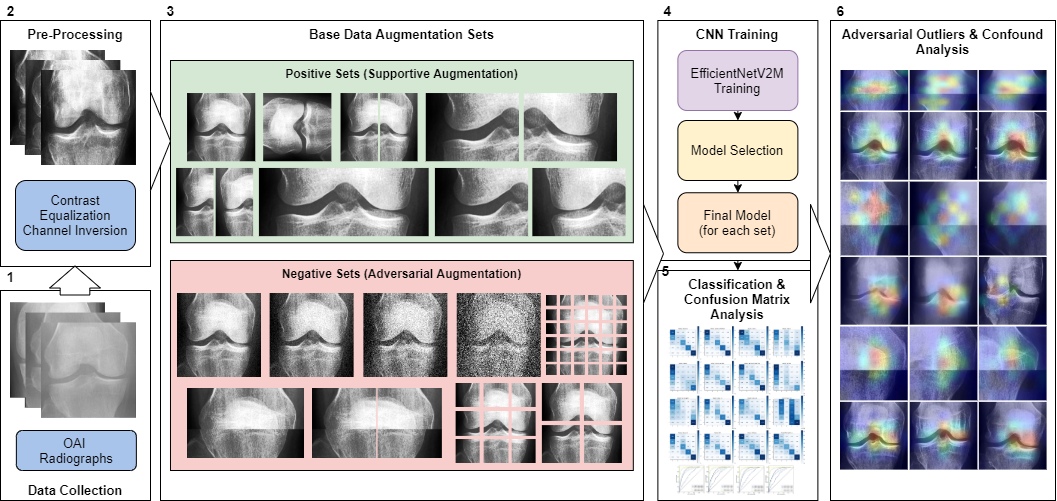}
\caption{The methodological pipeline for the study. Green represents positive/supportive augmentation, red signifies adversarial augmentations. Blue indicates data processing and purple signifies CNN training. Numeric markers indicate the order of operations.}
\label{flow}
\end{figure}

In this study, we present a comprehensive augmentation methodology for the classification of knee joint X-ray images sourced from the Osteoarthritis Initiative\cite{nevitt2006osteoarthritis}. Our approach is three-fold: data collection and preprocessing, image augmentation, and the application of a convolutional neural network (CNN) for classification. We utilized a dataset of 8260 images, graded via the Kellgren and Lawrence\cite{kellgren1957radiological} system, and subjected them to both positive/supportive and negative/adversarial augmentations. This was done to explore the benefits of artificial diversity during training and to challenge the classifier's resilience. The CNN model of choice was the EfficientNetV2-M\cite{tan2021efficientnetv2}, which was trained over 15 epochs with a dataset split into training, validation, and testing sets. To enhance the interpretability of our CNN model, we employed the Grad-CAM\cite{selvaraju2017grad} algorithm, providing visual insights into the decision-making process of the network. Our evaluation metrics included accuracy, precision, recall, and the F1 score, offering a wider view of the model's performance. Figure \ref{flow} illustrates the study's operation sequence using numeric markers. This section elaborates on each step in the order indicated by the numeric markers in the figure.

\subsection*{Data Collection}
\begin{figure}[!ht]
\centering
\includegraphics[scale=0.3]{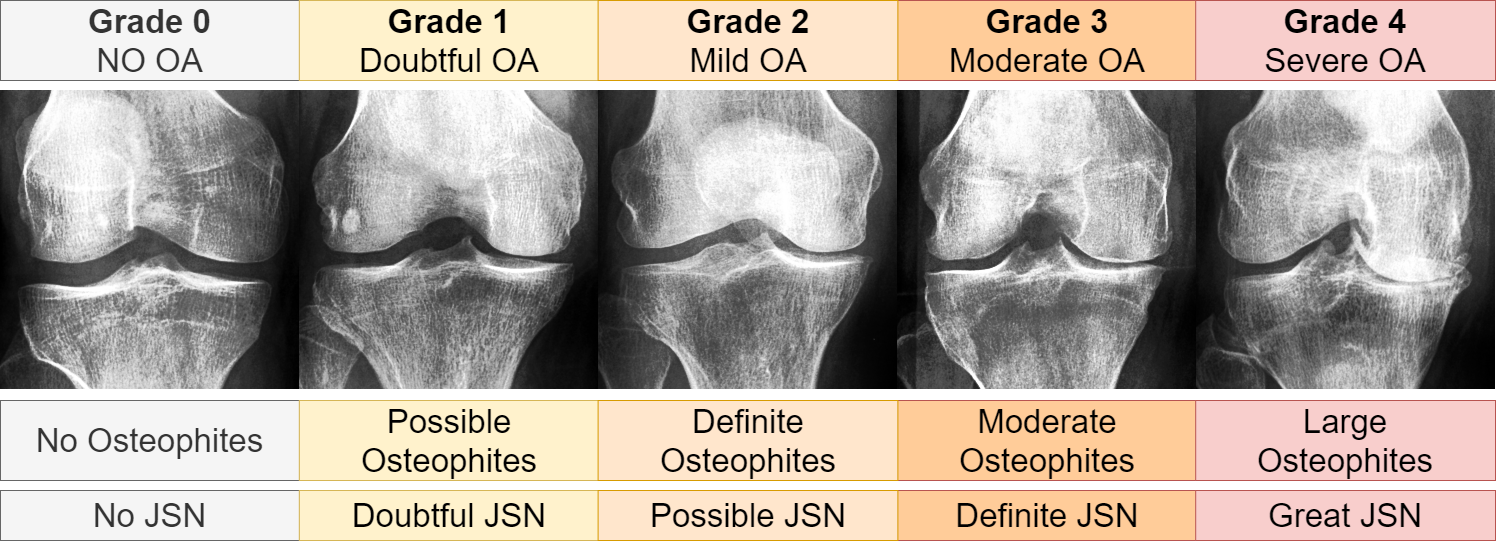}
\caption{Sample images showing various KL grades, ranging from 0 (no OA signs) to 4 (severe OA). From left to right, OA severity increases. Joint space narrowing, denoted as JSN.}
\label{grades}
\end{figure}

Our research utilized knee joint X-ray images from the Chen 2019 study\cite{chen2019fully}, sourced initially from the Osteoarthritis Initiative (OAI)\cite{nevitt2006osteoarthritis}. The OAI, a multi-center study focused on biomarkers for knee osteoarthritis, included 4796 participants aged 45 to 79. We employed the pre-processed primary cohort data from Chen 2019\cite{chen2019fully}, which had been subject to automatic knee joint detection, bounding, and zoom standardization to $0.14 \: \mathrm{mm/pixel}$. This led to 8260 images ($224 \times 224$ pixels) derived from 4130 X-rays containing both knee joints. The images were graded via the Kellgren and Lawrence (KL) system\cite{kellgren1957radiological}, as shown in Figure \ref{grades}. The KL grade distribution was as follows: 3253 images for Grade 0, 1495 for Grade 1, 2175 for Grade 2, 1086 for Grade 3, and 251 for Grade 4.

\subsection*{Image Pre-Processing}
\newcommand{\qvec}[1]{\textbf{\textit{#1}}}

We flipped each right knee joint image to mirror a left knee orientation. Then, we identified and inverted any negative channel images, resulting in 189 such alterations for KL01 and 77 for KL234. We then equalized the image histograms' contrast using equation \ref{eq:contr_norm}. In this equation, for a given grayscale image $\qvec{I}$ with dimensions of $m\times n $, we used the cumulative distribution function (cdf) and pixel value $v$ to obtain an equalized value $h(v)$ in the range $[0,255]$. Here, $\mathrm{cdf}_{\mathrm{min}}$ represents a non-zero minimum value of the image's cumulative distribution, while $m\times n$ signifies the total number of pixels.

\begin{equation}\label{eq:contr_norm}
h(v)=255\frac{\mathrm{cdf}(v)-\mathrm{cdf}_{\mathrm{min}}}{{(mn)-\mathrm{cdf}_{\mathrm{min}}}}
\end{equation}

\subsection*{Base Data Augmentation Sets}

\begin{figure}[!ht]
\centering
\includegraphics[scale=0.3]{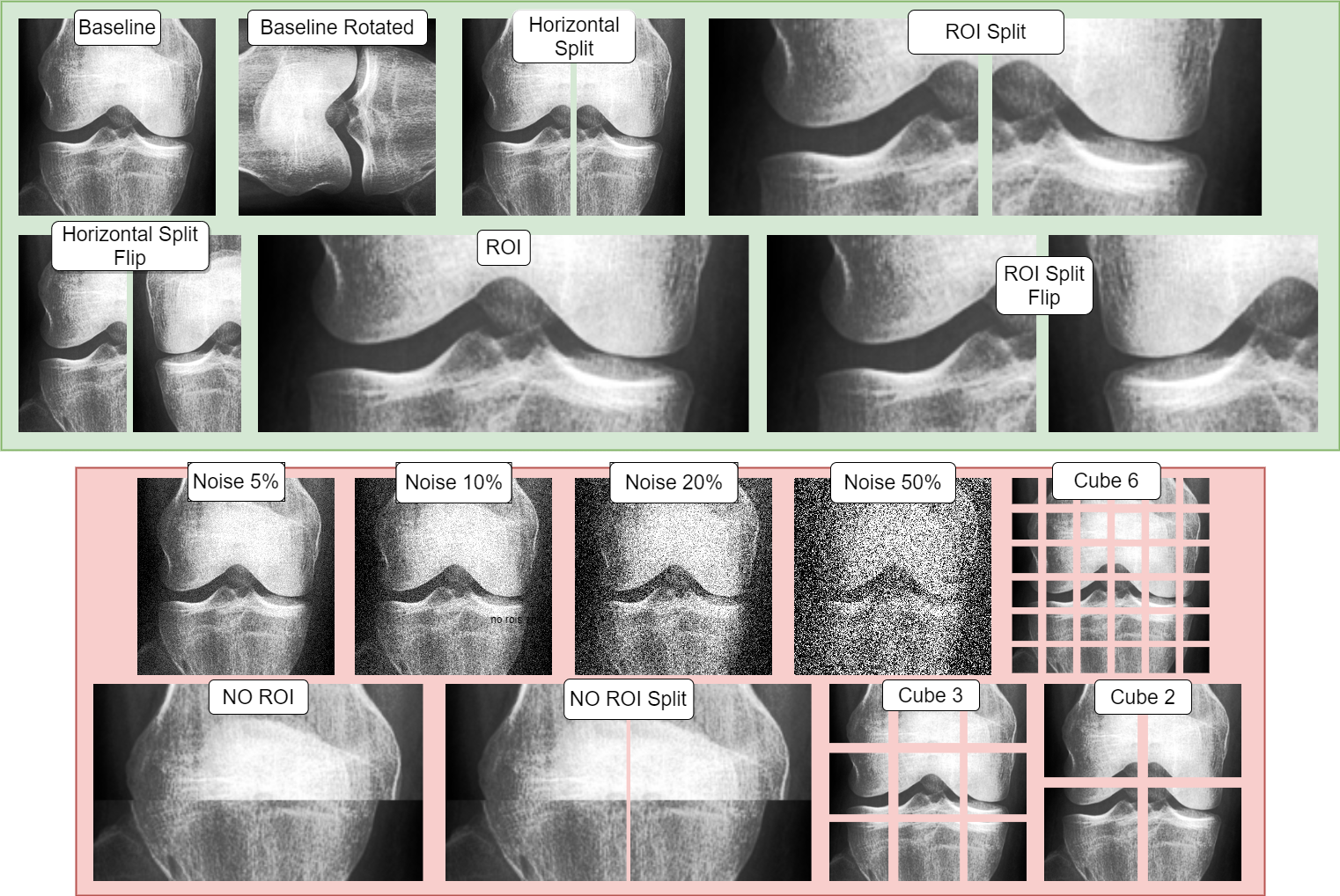}
\caption{Visualization of the study's base data augmentations: red indicates negative/adversarial augmentations and green shows positive/supportive augmentations. Each transformation is demonstrated on provided baseline image.}
\label{transforms}
\end{figure}

In our research, we divided our dataset into distinct splits and applied base data augmentations to each of these splits. We crafted two base data augmentation sets. The term 'base data' refers to enduring modifications made to all the data ('offline') before introducing any 'online' affine augmentations during the training phase (Table \ref{tab:afineaug}). The first augmentation set focused on positive or supportive modifications, exploring the potential benefits of incorporating artificial diversity during training. The second set, conversely, incorporated negative/adversarial augmentations intended to challenge the classifier. This was done to help pinpoint potential confounds in the classification task and test the model's resilience. Table \ref{tab:augmentations} showcases all conditions used, while figure \ref{transforms} visualizes the base augmentations made.

\begin{table}[!ht]
\centering
\begin{tabular}{|l|l|p{3cm}|}
\hline
\textbf{Augmentation Method} & \textbf{Description} & \textbf{Training Configuration} \\
\hline
Image Rotation & Implements a counter-clockwise rotation on images & Allows a rotation of up to 40 degrees \\
\hline
Width Shifting & Moves the image laterally along the x-axis & Allows a shift of up to 45 pixels along the x-axis \\
\hline
Height Shifting & Moves the image vertically along the y-axis & Allows a shift up to 45 pixels along the y-axis \\
\hline
Shearing & Distorts the image along the width or height axis & Implements a maximum shear angle of 0.2 degrees \\
\hline
Zooming & Modifies the image scale, zooming in or out within the frame & Enables a maximum of 20\% zoom \\
\hline
Horizontal Flipping & Creates a mirror image along the vertical axis & Implements flipping only along the horizontal axis \\
\hline
\end{tabular}
\caption{\label{tab:afineaug} Online augmentation approaches, occurring randomly during training}
\end{table}

\begin{table}[!ht]
\centering
\begin{tabular}{|l|l|p{7cm}|}
\hline
\textbf{Base Augmentation} & \textbf{Type} & \textbf{Description} \\
\hline
Baseline & Positive & Original image without any changes \\
\hline
Baseline Rotated & Positive & Original image rotated counter-clockwise \\
\hline
ROI\cite{wahyuningrum2019new} & Positive & Image segmented to highlight the Region of Interest \\
\hline
Horizontal Split 20\% & Positive & Image horizontally split in the middle with a 20\% overlap \\
\hline
Horizontal Split 20\% Flip & Positive & Image horizontally split in the middle with a 20\% overlap, and the second part flipped \\
\hline
Horizontal Split & Positive & Image horizontally split in the middle with a 5\% overlap \\
\hline
Horizontal Split Flip & Positive & Image horizontally split in the middle with a 5\% overlap, and the second part flipped \\
\hline
ROI Split\cite{tiulpin2018automatic} & Positive & ROI segmented image split in the middle with a 20\% overlap \\
\hline
ROI Split Flip & Positive & ROI segmented image split in the middle with a 20\% overlap, and the second part flipped \\
\hline
Noise 05 & Negative & 5\% Gaussian pixel noise added to the baseline \\
\hline
Noise 10 & Negative & 10\% Gaussian pixel noise added to the baseline \\
\hline
Noise 20 & Negative & 20\% Gaussian pixel noise added to the baseline \\
\hline
Noise 50 & Negative & 50\% Gaussian pixel noise added to the baseline \\
\hline
Cube 2 & Negative & Image divided into equidistant parts in a 2x2 grid \\
\hline
Cube 3\cite{wang2023transformer} & Negative & Image divided into equidistant parts in a 3x3 grid \\
\hline
Cube 6 & Negative & Image divided into equidistant parts in a 6x6 grid \\
\hline
No ROI & Negative & Concatenated upper and lower non-ROI parts of the image \\
\hline
No ROI Split & Negative & Concatenated upper and lower non-ROI parts of the image, split in the center \\
\hline
\end{tabular}
\caption{\label{tab:augmentations}Offline-additive base data augmentation sets, citations indicate the first known instance of the approach.}
\end{table}

\subsection*{Convolutional Neural Networks}
Convolutional neural networks (CNNs) \cite{lecun1995convolutional} are foundational in the recent deep learning revolution \cite{lecun2015deep}. CNNs are a type of neural network often used for computer vision. These neural networks employ the convolution operation between input and a filter-kernel. Filters slide across inputs to highlight features in a response known as a feature map. Various feature maps are combined to produce higher-level feature maps corresponding to higher-level concepts. Formally\cite{GoodBengCour16}, for an image $\qvec{I}$ of $u\times v$ dimensions and filter-kernel $\qvec{H}$ of $s\times t$ dimensions, we can obtain feature map $\qvec{G}$ by convolution across the two axes with kernel $\qvec{H}$ as:

\begin{equation}\label{eq:cnn}
\qvec{G}(u,v)=\sum_{s}\sum_{t}\qvec{I}(u,v)\qvec{H}(u-s,v-t)
\end{equation}

Typically, the feature map values are filtered with an activation function. The role of the activation function is to re-map the values across a given function. For example, a rectified linear unit activation function \cite{nair2010rectified} (ReLu) zeros-out negative values. Such an approach offers computational efficiency due to replacing redundant values with zero. For any feature map value $z$, the ReLu activation is defined as:

\begin{equation}\label{eq:relu}
f(z)= \max(0,z)
\end{equation}

In addition to the activation function operation, the max pooling operation is often used. Max pooling down-samples the convolution result such that cascades of max pooling and convolution would result in an ever-decreasing number of features. For image $\qvec{I}$ of $u\times v$ dimensions, the max pooled value ${g(u_I)}$ given dimension $u$ can be simply defined as follows:

\begin{equation}\label{eq:maxp}
g(u_I)= \lfloor\frac{{u_I}-r}{h}\rfloor+1
\end{equation}

Where $u_I$ is only dimension $u$ from image $\qvec{I}$, $r$ is the pooling window size, and $h$ is the stride value.

\subsection*{Convolutional Neural Network Architecture}

EfficientNet\cite{tan2019efficientnet}, a well-recognized deep learning model, employs compound scaling, balancing depth (number of layers), width (size of the layers), and resolution (size of the input image) in a structured manner. This scaling process is mathematically represented as:

\begin{equation}\label{eq:enet}
d = \alpha^{\phi} d_0, \quad w = \beta^{\phi} w_0, \quad r = \gamma^{\phi} r_0
\end{equation}

Here, $\alpha, \beta, \gamma$ are constants, $\phi$ is a user-defined coefficient, and $d_0, w_0, r_0$ represent depth, width, and resolution of the base model.

A vital component of EfficientNet is the MBConv block. It sequences transformations starting with a $1\times1$ convolution, a depth-wise convolution, a Squeeze-and-Excitation (SE) operation\cite{hu2018squeeze}, and another $1\times1$ convolution:

\begin{equation}\label{eq:mbconv_se}
T_{MB}(\qvec{I}) = \qvec{K}_2 \ast SE(\qvec{D} \ast (\qvec{K}_1 \ast \qvec{I}))
\end{equation}

In this formula, $\qvec{K}_1$ and $\qvec{K}_2$ are $1\times1$ convolutional filters, $\qvec{D}$ represents the depth-wise convolutional filter, and $\mathrm{SE}$ is the Squeeze-and-Excitation operation. EfficientNetV2\cite{tan2021efficientnetv2} extends the original model by incorporating a Fused-MBConv block, which combines the initial $1\times1$ and depth-wise convolutions into a single $3\times3$ convolution, followed by an SE operation and a final $1\times1$ convolution:

\begin{equation}\label{eq:fmbconv}
T_{FMB}(\qvec{I}) = \qvec{K}_{2} \ast (SE(\qvec{K}_{f} \ast \qvec{I}))
\end{equation}

\begin{figure} [!ht]
\centering
\includegraphics[scale=0.5]{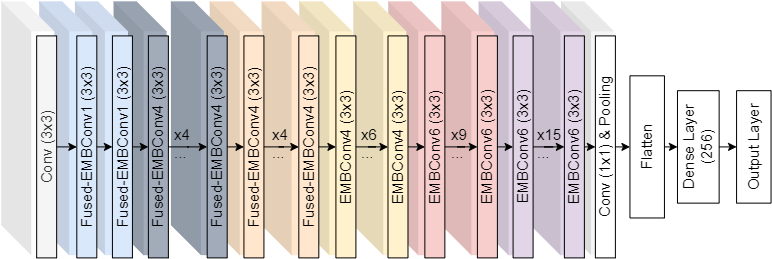}
\caption{EfficientNetV2 base architecture with our post-pooling modifications.}
\label{fig:effinet}
\end{figure}

Here, $\qvec{K}_{f}$ is the $3\times3$ convolutional filter combining the initial $1\times1$ and depth-wise convolutions, and $\qvec{K}_{2}$ is the final $1\times1$ convolutional filter. An activation follows each convolution and may include a skip connection. Figure \ref{fig:effinet} visualizes the base EfficientNetV2 model (B0) and our modifications.

Our study employed EfficientNetV2-M, enhancing it with flattening and 256-neuron dense layers. The training process was executed over 15 epochs using the Adam optimizer, with the dataset divided into Training (75\%), Validation (15\%), and Testing (15\%) sets (patient aware splits). The minimum validation loss determined early stopping. All CNN training and online affine augmentations were implemented with the open-source library Deep Fast Vision\cite{fabprezja_2023}. All offline affine augmentations are consistently applied to each image, but the degree to which they are applied is randomized within the specified ranges of each technique. These augmentations are directly incorporated and executed using the Keras library\cite{chollet2015keras} .

\subsection*{CNN Interpretability}
While the complexity of neural networks increases their capabilities, it also complicates the interpretation of their predictions\cite{prezja2023importance}. Due to this complexity, these systems are often deemed 'black boxes.' However, the Grad-CAM\cite{selvaraju2017grad} algorithm (based on the CAM\cite{Oquab_2015_CVPR} framework) helps reduce this 'black box' effect. At a high level, Grad-CAM is an algorithm that visualizes how a convolutional neural network makes its decisions. It creates what are known as "heat maps" or "activation maps" that highlight the areas in an input image that the model considers important for making its prediction. The Grad-CAM spatial activation map $M_{Grad-CAM}^{p}$ can be calculated using the $\mathrm{ReLU}$ activation function on the sum of neuron importance weights $b _{k}^{p}$ multiplied by feature maps $\Psi^{k}$ as shown below:

\begin{equation}\label{eq:gcam_combined}
M_{Grad-CAM}^{p} = \mathrm{ReLU}\left(\sum_{k} b {k}^{p} \Psi^{k}\right), \quad \text{where} \quad b {k}^{p} = \frac{1}{Z} \sum_{m} \sum_{n} \frac{\partial y ^{p}}{\partial \Psi _{mn}^{k}}
\end{equation}

In this equation, $b _{k}^{p}$ are the neuron importance weights of feature map $k$ for class $p$, $\frac{\partial y ^{p}}{\partial \Psi _{mn}^{k}}$ represents the partial derivative of the final layer prediction for class $p$ ($y ^{p}$) with respect to the last convolutional layer's $k$th feature map $\Psi _{mn}^{k}$. $Z$ is the total pixels, and $m,n$ are the indexes for each element within feature map $k$. $\Psi ^{k}$ is the feature map $k$ given by the last convolutional layer, spatially averaged. In our study, we extracted Grad-CAM activations from the layer immediately preceding the flattening operation.

\subsection*{Figures of Merit}
In evaluating the results of our experiment, we employ several key figures of merit to quantify the performance.

Accuracy is the proportion of true results (both true positives and true negatives) among the total number of cases examined. It can be calculated using the following equation:

\begin{equation}
\mathrm{Accuracy} = \frac{\mathrm{TP} + \mathrm{TN}}{\mathrm{TP} + \mathrm{TN} + \mathrm{FP} + \mathrm{FN}}
\end{equation}

Precision (also called positive predictive value) is the fraction of relevant instances among the retrieved instances. It is calculated as follows:

\begin{equation}
\mathrm{Precision} = \frac{\mathrm{TP}}{\mathrm{TP} + \mathrm{FP}}
\end{equation}

Recall (also known as sensitivity, hit rate, or true positive rate) is the fraction of the total amount of relevant instances that were actually retrieved. The equation for recall is:

\begin{equation}
\mathrm{Recall} = \frac{\mathrm{TP}}{\mathrm{TP} + \mathrm{FN}}
\end{equation}

The F1 score is the harmonic mean of precision and recall. It tries to find the balance between precision and recall. The F1 score can be calculated as follows:

\begin{equation}
\mathrm{F1} = 2 \frac{\mathrm{Precision} ~ \mathrm{Recall}}{\mathrm{Precision} + \mathrm{Recall}}
\end{equation}

In these formulas, TP is True Positives, TN is True Negatives, FP is False Positives, and FN is False Negatives.

\section*{Results}

\subsection*{Positive Augmentations}
In Table \ref{tab:positiveclrp}, the model with the best performance appeared to be the "Baseline Rotated" model, obtaining an Accuracy of 0.655, Precision of 0.621, Recall of 0.645, and an F1-Score of 0.618. The high accuracy indicated that this model was successful in correctly predicting the classification most of the time, while the substantial F1-Score, which is a harmonic mean of precision and recall, indicated a balanced high performance in both these areas. This suggested that the model could retrieve a high proportion of relevant instances (high recall), while ensuring the proportion of instances it claimed to be relevant were indeed relevant (high precision).

In contrast, the model with the lowest performance in the evaluation was the "Horizontal Split" model. With an Accuracy of 0.560, Precision of 0.497, Recall of 0.563, and an F1-Score of 0.501, this model consistently fell behind the other models across all performance metrics, indicating lower overall performance. It can be observed that there was a clear downward trend in performance metrics from the "Baseline Rotated" model to the "Horizontal Split" model. Interestingly, models using the "Flip" modification, such as "Horizontal Split Flip" and "ROI Split Flip", tended to have a lower performance than their non-flip counterparts. The only exception to this was the "Horizontal Split 20\% Flip" model, which slightly outperformed the "ROI Split" and "Horizontal Split" models, suggesting that the impact of the "Flip" modification could be influenced by broader image overalp (0.20 in that case).

\begin{table}[!ht]
\centering
\begin{tabular}{|l|l|l|l|l|}
\hline
\textbf{Model Name}            & \textbf{Accuracy} & \textbf{Precision} & \textbf{Recall} & \textbf{F1-Score} \\ \hline
Baseline Rotated               & 0.655             & 0.621              & 0.645           & 0.618             \\ \hline
Baseline                       & 0.644             & 0.621              & 0.651           & 0.627             \\ \hline
ROI                            & 0.623             & 0.614              & 0.643           & 0.616             \\ \hline
Horizontal Split 20\%          & 0.595             & 0.597              & 0.611           & 0.591             \\ \hline
Horizontal Split Flip          & 0.581             & 0.518              & 0.577           & 0.531             \\ \hline
ROI Split Flip                 & 0.583             & 0.577              & 0.595           & 0.579             \\ \hline
Horizontal Split 20\% Flip     & 0.577             & 0.578              & 0.591           & 0.570             \\ \hline
ROI Split                      & 0.570             & 0.533              & 0.579           & 0.546             \\ \hline
Horizontal Split               & 0.560             & 0.497              & 0.563           & 0.501             \\ \hline
\end{tabular}
\caption{Performance metrics of the model on the test set using positive/supportive base augmentations.\label{tab:positiveclrp}}
\end{table}

\begin{figure}[!ht]
\centering
\includegraphics[scale=0.20]{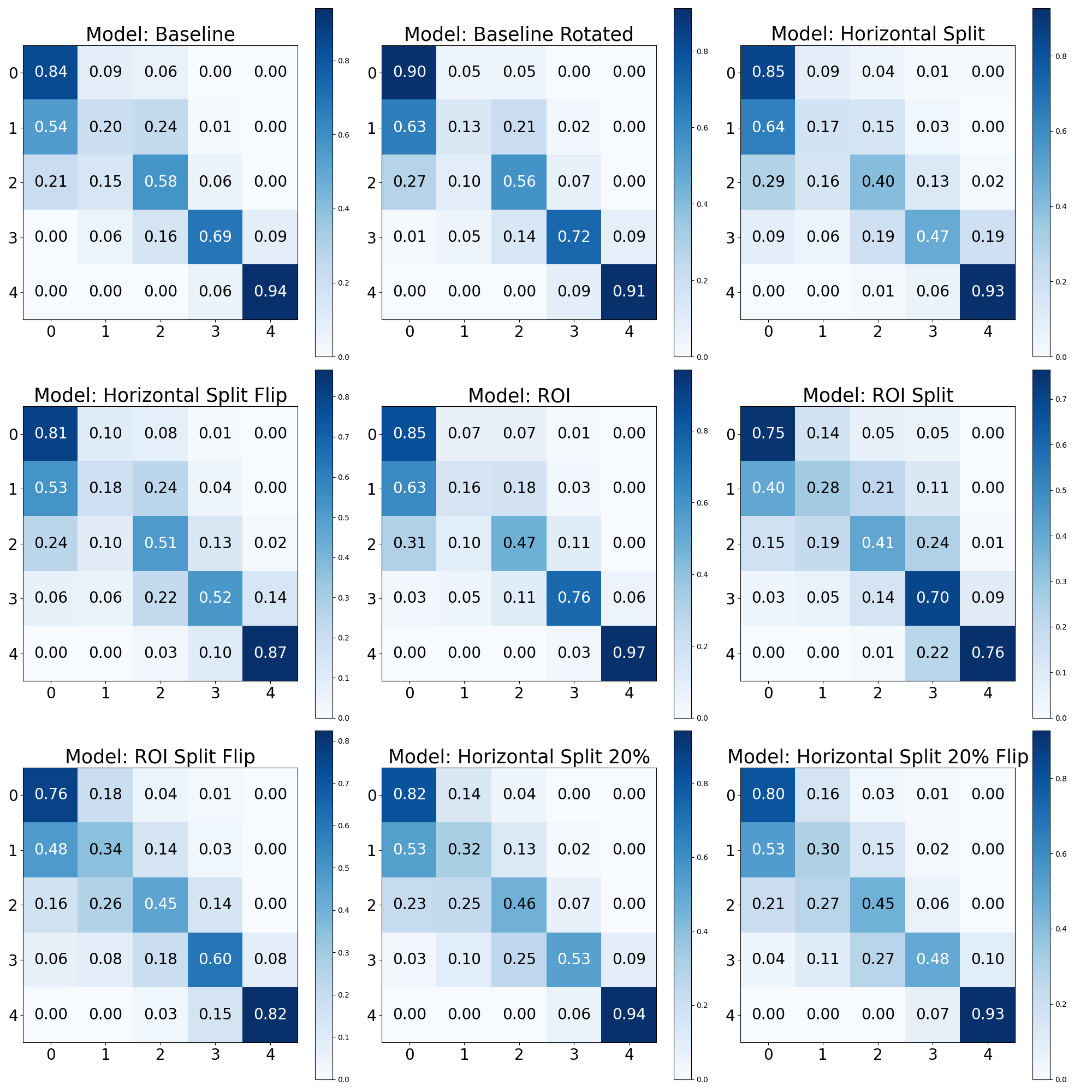}
\caption{Confusion matrices for the test set using positive/supportive base data augmentations.}
\label{positiveconv}
\end{figure}

The provided confusion matrix \ref{positiveconv} revealed the performance of nine distinct models: Baseline, Baseline Rotated, Horizontal Split, Horizontal Split Flip, ROI, ROI Split, ROI Split Flip, Horizontal Split 20\%, and Horizontal Split 20\% Flip. The Baseline model performed well for KL0 and KL4 but encountered difficulties with the intermediate classes. This issue was partly alleviated in the Baseline Rotated model for KL0. The Horizontal Split and Flip models demonstrated variability in performance across classes, with the Flip version slightly improving the accuracy for KL2 and KL3. The Region of Interest (ROI) models exhibited improvements for the intermediate classes, especially KL2. The ROI Split and Flip models offered a more balanced performance across classes, particularly for KL0 and KL1. Lastly, the Horizontal Split 20\% and its Flip variant showed high misclassification rates between KL0 and KL1, although the Flip version brought some improvement. However, KL4 was well classified across all models, suggesting distinct features that differentiated it from other classes.

Models with the "Flip" modification seemed to have a more evenly distributed confusion matrix, indicating a more balanced prediction across different classes. However, this did not always result in overall higher performance, as seen in the "Horizontal Split Flip" model. The "Baseline Rotated" model seemed to perform well for the first and last class, but its performance decreased notably for the other classes. This behavior was shared across models, where models often performed better for the first and last class. The "ROI" and "ROI Split" models presented a similar pattern, with typical performance in the first and last classes.

\begin{figure}[!ht]
\centering
\includegraphics[scale=0.38]{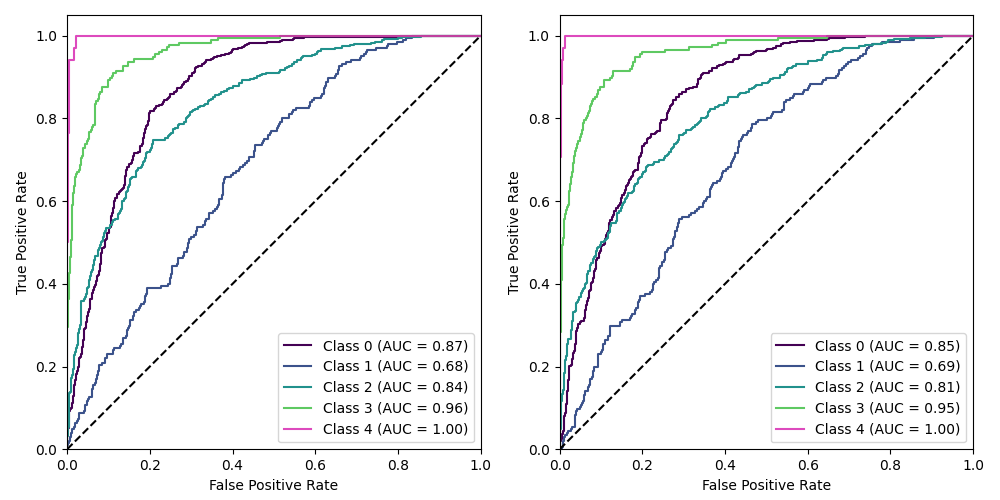}
\caption{Using a one-vs-all scheme, ROC curves are shown for the Baseline condition (left) and the ROI condition (right).}
\label{positiveroc}
\end{figure}

Reviewing the results from the different model iterations, it was quite unexpected to find that the ROI (Region of Interest) models did not manage to outperform the Baseline models. Given that the Baseline models leveraged the entire image and the ROI models focused on the specific region expected to contain more relevant information for the task, it was anticipated that the ROI models would perform better. Figure \ref{positiveroc} showcases ROC curves for the ROI model against the baseline.

Upon evaluation of AUCs in the one-vs-all scheme (Figure \ref{positiveroc}), it was observed that Classes 0, 2, and 3 had marginally higher AUC scores in the Baseline Model compared to the ROI Model. Specifically, the AUC for Class 0 was 0.87 in the Baseline Model versus 0.85 in the ROI Model, suggesting that the Baseline Model was slightly more successful in distinguishing between positive and negative instances for this class. Similarly, for Class 2, the Baseline Model had an AUC of 0.84 compared to 0.81 in the ROI Model, and for Class 3, the AUC was 0.96 in the Baseline Model versus 0.95 in the ROI Model. In contrast, for Class 1, the ROI Model outperformed the Baseline Model, albeit slightly, with an AUC of 0.69 against 0.68. For Class 4, both models performed impeccably, achieving a perfect AUC score of 1.00, demonstrating their ability to distinguish instances of this class perfectly.

In summary, while the performance of both models was similar for all classes, the Baseline Model showed a slight edge in Classes 0, 2, and 3. The ROI Model only performed marginally better in Class 1, and both models were equally successful in Class 4. Despite these differences in AUC values, it's noted that the curvature along the axis was similar between the two models. This suggestsed that the trade-off between sensitivity and specificity (true positive rate and false positive rate) was similar for both models across different decision thresholds. This similarity in shape indicated that both models had similar performance trade-offs, even if the absolute performance (as measured by AUC) varied slightly.

\subsection*{Negative (Adversarial) Augmentations}
In table \ref{tab:negativeclr}, we find a comparison of several adversarial augmentation models based on metrics including Accuracy, Precision, Recall, and F1-Score. The model with noise level 05 had the highest performance across all metrics. As noise increased (i.e., Noise 10 and Noise 20), a corresponding decrease was observed in all performance metrics. This suggested that lower levels of noise improved the model's ability to generalize. In comparison, higher noise levels degraded the performance, likely due to interference with essential radiograph features. The models using cube techniques also showed varying levels of performance. Cube 2, for instance, performed better than Cube 3 and Cube 6 in all aspects. This could imply a potential optimal size or representation for the cube that best captured critical information. The models with no Region of Interest (ROI) performed poorly compared to others. The Noise 50 model recorded the lowest performance.

\begin{table}[!ht]
\centering
\begin{tabular}{|l|l|l|l|l|}
\hline
\textbf{Model Name} & \textbf{Accuracy} & \textbf{Precision} & \textbf{Recall} & \textbf{F1-Score} \\ \hline
Noise 05            & 0.619             & 0.592              & 0.616           & 0.591             \\ \hline
Noise 10            & 0.591             & 0.572              & 0.601           & 0.574             \\ \hline
Noise 20            & 0.478             & 0.489              & 0.544           & 0.495             \\ \hline
Cube 2              & 0.483             & 0.469              & 0.502           & 0.458             \\ \hline
No ROI              & 0.427             & 0.315              & 0.343           & 0.289             \\ \hline
Cube 3              & 0.407             & 0.347              & 0.406           & 0.299             \\ \hline
Cube 6              & 0.359             & 0.287              & 0.333           & 0.272             \\ \hline
No ROI Split        & 0.349             & 0.294              & 0.371           & 0.263             \\ \hline
Noise 50            & 0.186             & 0.262              & 0.291           & 0.174             \\ \hline
\end{tabular}
\caption{Performance metrics of the model on the test set using negative/adversarial base augmentations.\label{tab:negativeclr}}
\end{table}

\begin{figure}[!ht]
\centering
\includegraphics[scale=0.20]{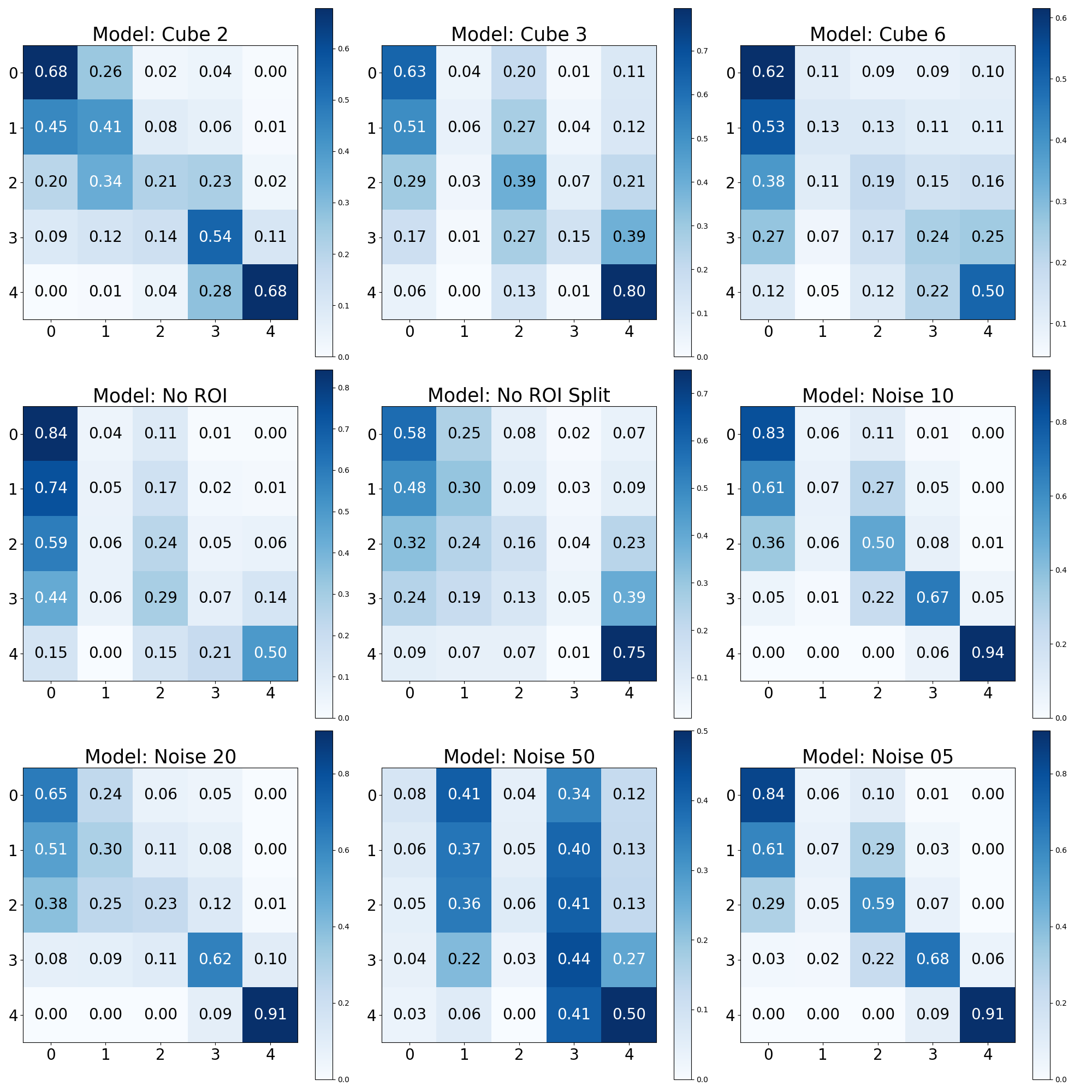}
\caption{Confusion matrices for the test set using negative/adversarial base data augmentations.}
\label{negativeconf}
\end{figure}

In figure \ref{negativeconf}, the Cube 3 model performed well when identifying KL0; however, as the Kellgren-Lawrence (KL) grades increased, this performance gradually diminished, culminating in a notable difficulty when classifying KL4. This suggested that while the model could easily distinguish KL0 from other classes, the higher grades posed more of a challenge. In contrast, the Cube 2 model showed a more even performance across all KL grades, with a gradual decrease in accuracy from KL0 to KL4. While the model also performed best on KL0 and worst on KL4, it demonstrated a more balanced confusion across different classes. The Cube 6 model continued the trend observed in Cube 3 and Cube 2, struggling with higher KL grades and decreasing performance from KL0 to KL4. Interestingly, it confused KL0 with KL1 and KL3 more than Cube 3 and Cube 2, which indicated its difficulty differentiating between these classes.

Shifting to the Noise models, Noise 10 and Noise 05 displayed interesting behavior. They were fairly accurate when classifying the extreme KL grades (KL0 and KL4), but encountered difficulties with the intermediate classes. The Noise 20 model, like Cube 2, demonstrated an evenly distributed performance across the grades, but it performed slightly better on class KL4 than Cube 2. This could point to a better ability to differentiate between the features of the KL4 grade. The Noise 50 model was unique in its performance, showing a high confusion rate, particularly between KL0 and KL1, and KL1 and KL3. It also struggled with KL0, KL1, and KL2, yet performed reasonably well on KL4, indicating that this model found the lower and intermediate KL grades more challenging.

Lastly, the No ROI model showed exceptional performance when classifying KL0, but it struggled to differentiate KL0 from KL2, and performed notably poorly on KL3. Similarly, the No ROI Split model showed a distinct performance pattern. It was particularly adept at identifying KL0 and KL4.

In the No ROI model, the high score for the KL0 class (0.84) indicated that the model effectively identified patterns associated with KL0. However, there is an absence of the primary region of interest. Similarly, the No ROI Split model achieved a surprisingly high score of 0.75 for the KL4 class. This suggested that these models were identifying other image features unrelated to the knee joint to make the classification decisions for these particular grades.

\begin{figure}[!ht]
\centering
\includegraphics[scale=0.38]{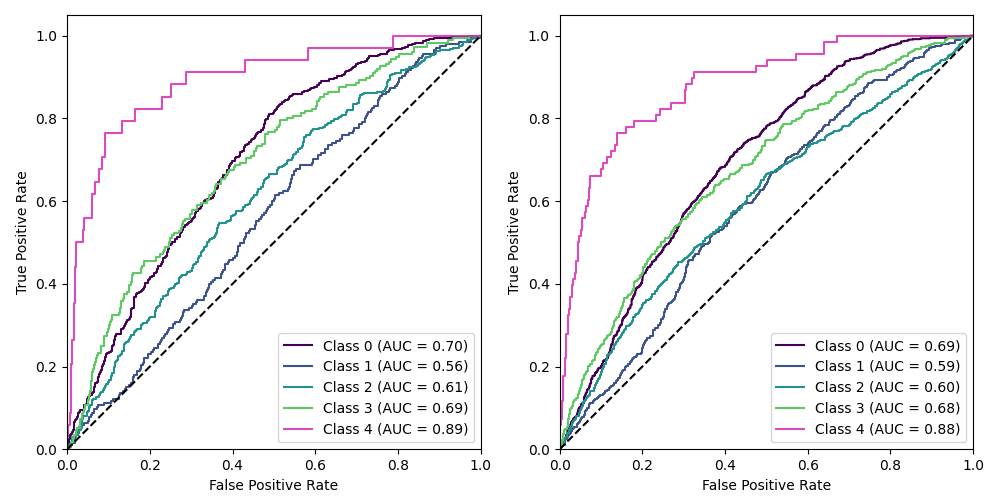}
\caption{Using a one-vs-all scheme, ROC curves are shown for the NO ROI condition (left) and the NO ROI Split condition (right).}
\label{negativeroc}
\end{figure}

Figure \ref{negativeroc} illustrates the Receiver Operating Characteristic (ROC) curves for both the "no ROI" and the "no ROI split" configurations. Notably, the model's performance appeared virtually indistinguishable across these settings when employing a one-versus-all scheme. However, we must also pay attention to the markedly high Area Under Curve (AUC) values for KL 0 (> 0.70) and KL 4 (> 0.88). These elevated values, alongside the accompanying confusion matrices, underscored the prevalence of potential confounding regions within these images. Remarkably, these potential confounding regions enabled a level of classification precision that is both significant and surprising, particularly given the absence of a region of interest, such as the entire knee joint. We further investigated these outlier results in the next section.

\subsection*{Adversarial Outliers}
In this section, we extend our results corresponding to the identified outlier classifications - specifically, "NO ROI" and "NO ROI Split". To enhance the depth and clarity of our analysis, we juxtapose these outcomes with the baseline results, enabling a more thorough comparative evaluation.

\begin{figure}[!ht]
\centering
\includegraphics[scale=0.23]{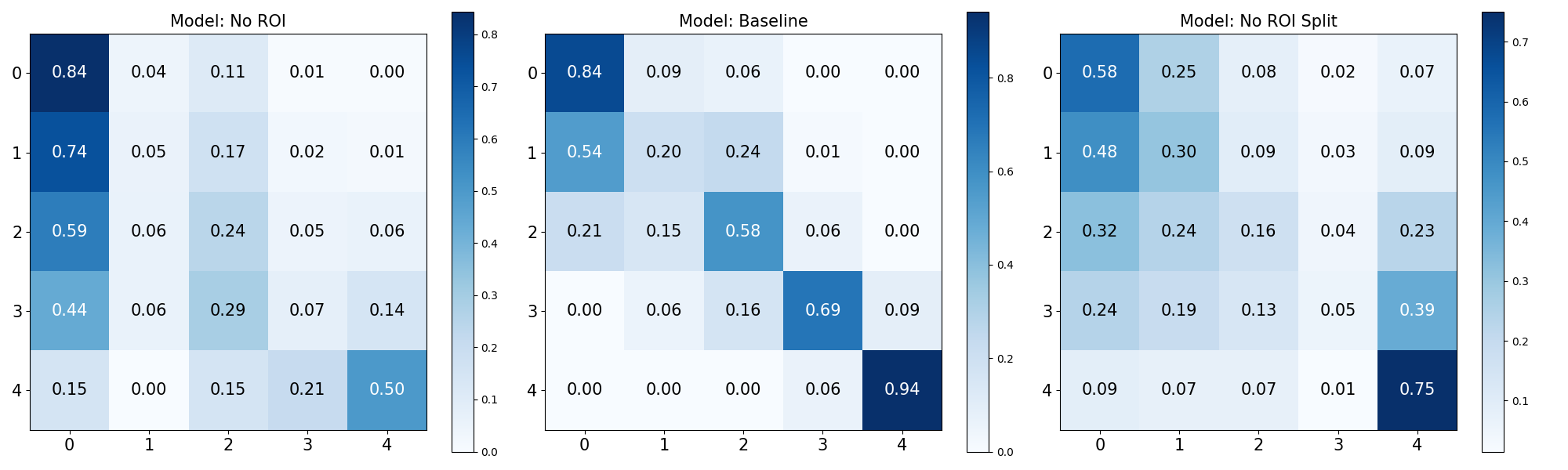}
\caption{Confusion matrices for the test set using no ROI, baseline, and no ROI split base data augmentations.}
\label{outlierconf}
\end{figure}

In Figure \ref{outlierconf}, we noticed identical scores from the 'Baseline' and 'No ROI' models in the case of KL0, which proposed either that the absence of a large region of interest (ROI) does not affect KL0 classification or that true class region confounds are visible. Moreover, the 'No ROI Split' model demonstrated performance remarkably similar to the 'Baseline' model for KL4. Although it did not reach complete alignment with the 'Baseline,' its relative success hinted at similar causes as observed in KL0. Most interestingly, we observed a clear performance boost for KL1 in the 'NO ROI Split' model. This class is historically the most significant challenge for classification in Knee-Osteoarthritis. Remarkably, this score was the highest individual KL1 score across all examined models of this study.

\begin{figure}[!ht]
\centering
\includegraphics[scale=0.172]{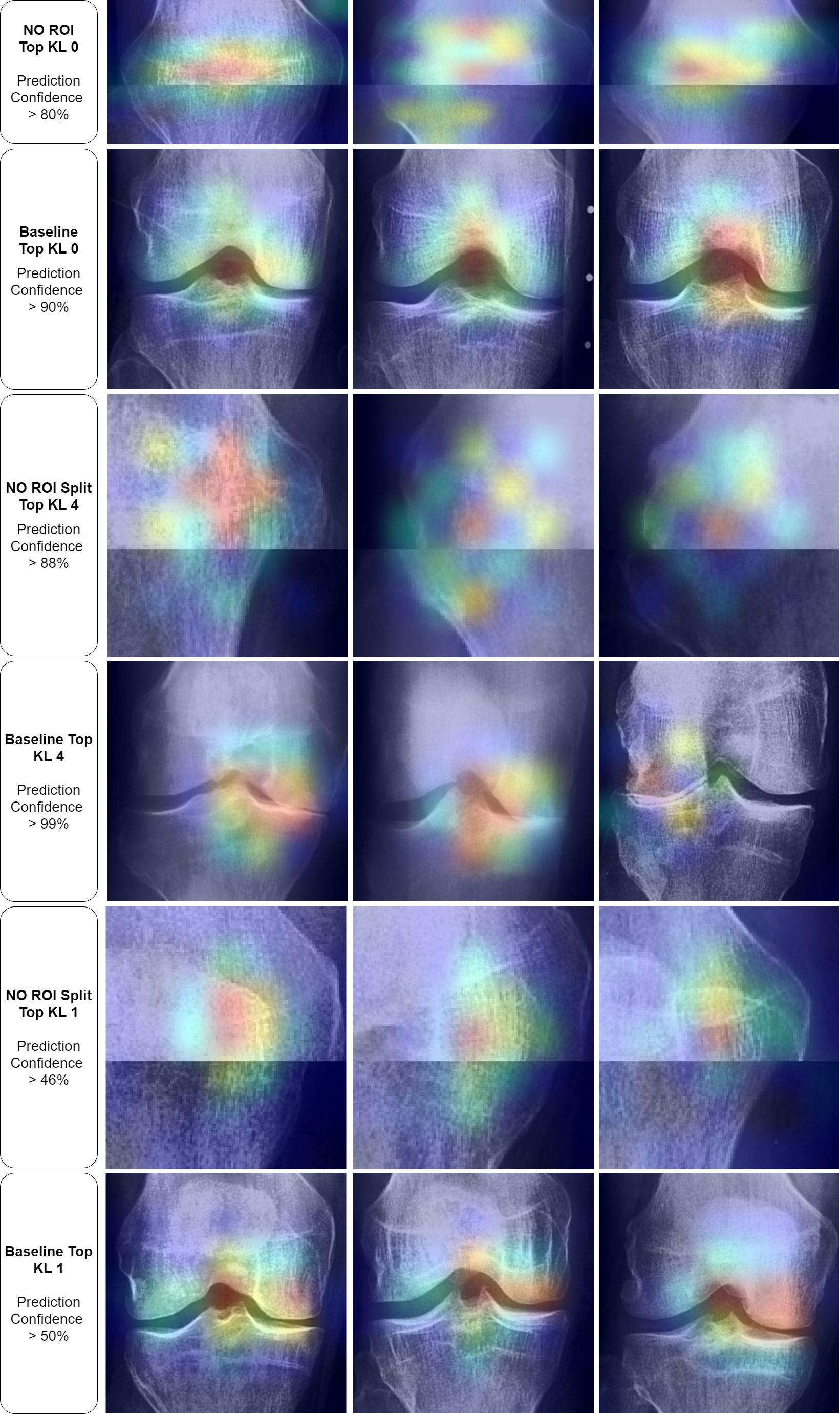}
\caption{Grad-CAM displays for outlier conditions with baseline comparisons. 'Confidence' denotes the output layer's value on the correct class.}
\label{gradcam}
\end{figure}

To delve deeper into these paradoxical results, we applied the grad-CAM technique to the top examples from each of the previously mentioned classes. As shown in Figure \ref{gradcam}, in the first row (no ROI KL0), we observed activations focused on texture and potential outlines of the patella. On the other hand, the second row (Baseline KL0) displayed control images that distinctly highlighted the knee joint. However, it is essential to note the broad spread of activation extending across and above the knee joint. Observations in the third row (No ROI Split KL4) revealed unclear patterns, primarily centered around what seemed to be a focus on wear-related texture. Despite the ambiguity, the controls (row 4) highlighted the knee joint, albeit with significantly less spread than the KL0 control (second row). Examining the KL1 focus in the first image of the last set (fourth row) revealed what appeared to be a part of the patella outline. In contrast, the other two images (fourth row) depicted a non-specific texture focus. Finally, all controls (KL0, KL4, KL1) presented a wide activation range that extended slightly over the knee joint. Overall, our observations suggested that the baseline models for KL0 tend to relied not only on the knee joint but also incorporated broader areas for their classifications. This starkly contrasted the KL1 baseline models, which seemed to concentrate more on the knee joint.

\section*{Discussion}
Our analysis has highlighted the marked effectiveness of certain positive data augmentations in improving model performance. Specifically, the 'Baseline Rotated' model showed the highest performance. Incorporating rotation into our baseline model could have increased its robustness to orientation flips in the images, contributing to its superior performance. Furthermore, the confusion matrix analysis demonstrated an excellent performance for the KL0 and KL4 grades, a result that may be associated with the distinct radiographic features of these classes. Conversely, the 'Horizontal Split' model, which divided the image into two parts along the horizontal axis, performed the worst across all the considered metrics. This could be because this approach might eliminate or distort crucial radiographic features, thereby reducing the model's ability to classify the images accurately. Notably, the results contradicted our initial expectation that the ROI models would outperform the baseline models, given the assumption that focusing on specific regions containing more relevant information would increase performance. The results, however, indicated that models which utilize the entire image data might have a slight edge in performance over those that focus on a particular ROI, suggesting either that potentially important information outside the ROI might be missed or that confounds are integrated to inflate the performance.

While data augmentation techniques have been widely adopted in deep learning, studies specifically investigating their effects in the medical imaging domain remain sparse. Often, the choice of augmentation techniques relies heavily on informal recommendations or generic best practices that aren't always tailored to medical images' unique challenges and characteristics\cite{goceri2023medical}. This lack of systematic exploration can lead to suboptimal model performance or even introduce biases. Against this backdrop, Goceri\cite{goceri2023medical} and Hussain, et al.\cite{hussain2017differential} studies stand out as notable exceptions that delve deep into the intricacies of various additive augmentation methods and their impact on model performance in medical imaging tasks.

The study by Goceri\cite{goceri2023medical} spanning different medical imaging domains such as lung CT, mammography, and brain MR images observed distinct patterns in additive augmentation effectiveness. For lung CT images, translating and shearing produced the highest accuracy of 0.857, whereas a mere translation yielded the lowest at 0.610. In mammography images, the combination of translation, shearing, and clockwise rotation was most effective with an accuracy of 0.833, while adding 'salt-and-pepper' noise and shearing underperformed, achieving only 0.667. For brain MR images, the same combination of translation, shearing, and clockwise rotation outperformed other methods with an accuracy of 0.882, adding 'salt-and-pepper' noise and shearing, showing the lowest accuracy at 0.624. Another investigation by Hussain, et al\cite{hussain2017differential} explored different mammography additive augmentation techniques, producing varied results. Notably, the Shear augmentation achieved the highest training accuracy of 0.891 and a validation accuracy of 0.879. Conversely, Noise augmentation was the least effective, with training and validation accuracies of 0.625 and 0.660, respectively. Augmentations such as Gaussian Filter, Rotate, and Scale also demonstrated high accuracy in training and validation phases. By comparing our results, those of Goceri, and the findings from Hussain, et al, it becomes evident that while some augmentation methods consistently show effectiveness across studies, the efficacy can vary based on domain specificity and dataset nuances. Our results, especially those pertaining to the 'Baseline Rotated' model, suggest that certain augmentations, such as rotation, might have unique advantages in the context of KOA.

Negative augmentations, in the form of adversarial attacks, were explored in our study. It was observed that as the noise level increased, the models' performance deteriorated, suggesting that the introduction of excessive noise could disrupt the discernment of relevant features within the images. Interestingly, the models lacking a Region of Interest (ROI) performed poorly overall. However, these models did exhibit exceptionally high performance for specific KL grades, such as KL0 and KL4, which may indicate the presence of confounding variables that the model is leveraging to make its predictions. The results for the "No ROI" and "No ROI Split" models were particularly intriguing. Despite the absence of a region of interest, the high-performance scores achieved by these models for specific KL grades suggest that these models might be identifying other image features unrelated directly to knee joint osteoarthritis to make the classification decisions. In the case of KL0, the identical scores from the 'Baseline' and 'No ROI' models proposed either that the absence of a region of interest (ROI) may not affect KL0 classification or that true class confounds are visible. For KL4, the 'No ROI Split' model demonstrated performance remarkably similar to the 'Baseline' model, hinting at similar influences. The most notable result, however, was the clear performance boost for KL1 in the 'NO ROI Split' model. This class is historically challenging to classify in Knee-Osteoarthritis studies, which makes this finding of particular interest.

Our Grad-CAM visualization analysis revealed insights into the potential confounding regions that might affect our models' decision-making processes. Notably, in the absence of a designated region of interest (ROI) for KL0, the models show a tendency towards the texture and contours of the patella. Interestingly, this pattern shifts with the baseline KL0 models, where the joint and its eminence are distinctly highlighted. However, the spread of activation that extends broadly across and above the knee joint suggests the model might be considering features beyond the knee joint for classification. In the No ROI Split KL4. the model appeared to be using general wear-and-tear texture indications for their classifications, which may not be directly related to disease progression but rather simply the participant's age. Finally, in the KL1 category, the models oscillate between specific and non-specific textures, further underscoring potential confounding regions.

Our best-performing set involved rotation, which may be partly attributed to the rotated orientation of the knee joint along the vertical axis of the radiograph. In this configuration, the convolution operation repeatedly encounters relevant features as it slides across the image, potentially leading to a more condensed and effective feature maps. This is in contrast to a non-rotated radiograph, where the knee joint occupies only a single vertical segment of the image. In this latter case, the convolution would likely traverse the entire joint just once or twice, depending on the receptive field, making feature extraction potentially slightly less effective.

Our findings regarding pixel noise underscore an essential consideration in raw radiograph data, which may contain parts of suboptimal quality with substantial pixel noise. These observations are in line with similar findings reported in other studies. The sensitivity to noise is especially evident for the early stages of the condition, as even a minimal noise level of $5 \: \%$ led to a decline in performance. This raises the question of whether including low-quality radiographs might be more detrimental than beneficial. This observation suggests a potential direction for future research to establish a guideline for acceptable noise levels. An alternative approach could be incorporating blur effects in the augmentation process to counteract noise-related challenges. However, the efficacy of such methods is beyond the scope of this study and warrants further investigation.

\section*{Conclusion}

In this study, we evaluated the effectiveness of various data augmentation techniques to enhance model performance in knee-joint osteoarthritis classification. These findings have potential implications for future work in this area, particularly for improving the robustness and accuracy of deep-learning models in medical image analysis. However, our results also highlight the need to carefully consider potential confounding regions to ensure that the models primarily base their predictions on relevant features. To facilitate further analyses, we provide open access to all data, trained models, and an extensive set of the top 20 Grad-CAM images, ranked by prediction confidence. This information is available in our data availability section.

\bibliography{main}

\section*{Acknowledgements}
The authors extend their sincere gratitude to Kimmo Riihiaho, Rodion Enkel, Leevi Lind and Suvi Lahtinen.

\section*{Data Availability}
All trained models, data, and grad-cam images from the current study are available the Google drive repository.

\section*{Author contributions statement}
Conceptualization: All Authors;
Methodology: F. P.; 
Data Curation: All authors;
Writing – review \& editing: All authors.

\section*{Additional information}
 \textbf{Competing interests}
 All authors declare that they have no conflicts of interest.
\end{document}